\documentclass[twoside]{article}
\usepackage[latin1]{inputenc}
\usepackage{t1enc}
\usepackage{a4}
\usepackage{tabularx}
\usepackage{epsf}

\def\bref{\vspace{4pt}\noindent\hangindent=10mm}

\begin{document}

\setcounter{figure}{0}
\setcounter{section}{0}
\setcounter{equation}{0}

\begin{center}
{\Large\bf
Solar System Formation}\\[0.7cm]

Aur\'elien {\sc Crida}\\[0.17cm]
Institut f\"ur Astronomie \& Astrophysik, Universit\"at T\"ubingen \\
Abt. Computational Physics,\\
Auf der Morgenstelle 10\\
D-72076 T\"ubingen\\
GERMANY\\

{\tt crida@tat.physik.uni-tuebingen.de}
\end{center}

\vspace{0.5cm}

\begin{abstract}
\noindent{\it
This review concerns the formation of bodies of the Solar System. Three major changes in our understanding of the early history of the Solar System are presented.

{\emph 1) Early differentiation:}
A few recent results support the idea that protoplanet formation and differentiation occurred partly simultaneously than CAI formation (which is generally accepted as time 0 of the Solar System). First, some iron meteorites, eucrites, and angrites older than the chondrules or even than the CAI have been found. Second, iron meteorites could be debris of early disrupted differentiated planetesimals, scattered from the terrestrial planet region to the Main Belt. Finally, chondrules contain fragments of planetesimal material.

{\emph 2) Earth and Moon:}
An equilibration mechanism explains the identical Oxygen isotopic composition of the Earth and the Moon. In addition, it has been shown that the Earth and the Moon mantles have the same $^{182}$W anomaly, in contrast to what was believed before. Consequently, the Moon forming impact should have occurred after the extinction of the $^{182}$Hf radioactivity, about $60$~Myr after Solar System formation. This new datation is in agreement with new N-body numerical simulations of the last phase of terrestrial planets formation, in which giant impacts occur during about $100$~Myr.

{\emph 3) Giant planets and Nice model:}
The migration of the giant planets in the protoplanetary disc can be prevented if the planets are in resonance, close to each other. In 2005, Tsiganis, Gomes, Morbidelli, and Levison have proposed a model in which the 4 outer planets of the Solar System were in a compact configuration after the dissipation of gaseous disc. A few hundred million years later, a global instability drives the planets on their present orbits, producing the Late Heavy Bombardment. In the frame of this so called ``Nice-model'', a lot of characteristics of our Solar System can be explained (distribution of the Trojans and Hildas asteroids, Kuiper belt structure, irregular satellites of the giant planets\ldots).
}
\end{abstract}

\section{Introduction}

The Solar System is the most well known astronomical object. A considerable amount of well constrained properties have to be explained, and a lot of observations are available. This makes the study of its formation easy and difficult at the same time, exciting in any case. This review on the formation of the Solar System focuses on recent results that are particular to \emph{our} Solar System. General works on planet formation will be ignored. In the last few years, new results have deeply changed our idea of three aspects of the Solar System formation, leading to new ways of thinking the early history of the planets. These three new scenarii have no connection with each other, but all three of them represent a fascinating improvement in our understanding of the formation of the Solar System. Therefore, this review is divided in three independent chapters, organised in order of increasing size and duration of the processes, from the smaller to the larger scales.

In section~\ref{met}, we focus on the first two million years of the Solar System and on the small bodies. Recent observations, datations and modelisations of meteorites, are reviewed. New results, obtained with isotopic datation, numerical simulations, and petrographic analysis, suggest that the standard idea of planet formation from chondrules as elementary bricks should be revised. On the contrary, differentiation of small bodies appears to have taken place before the formation of chondrules. A new scenario is presented, in which these works take place.

In section~\ref{EM}, we shift to terrestrial planets size, and several dozens of millions of years\,: new results on the Moon and the date of the Moon forming impact are presented. The isotopic composition of our satellite is explained in Oxygen and Hafnium, and its age is revised. N-body simulations as well as isotopic datation suggest a Moon forming impact later than previously expected, 60 to 100 million years after Solar System formation instead of about 40.

In section~\ref{Nice}, the dynamics of the giant planets over a billion of years is discussed. First, it has been shown that their migration in the gaseous protoplanetary disc can be prevented if they interact with each other, in some conditions. Then, after the protoplanetary disc is dissipated, the ``Nice model'' can be applied. Slowly perturbed by an outer disc of planetesimals during hundreds of millions of years, the giant planets suffer a global instability that drives them to their present orbits and causes the Late Heavy Bombardment of the terrestrial planets. During the instability, the Trojans of Jupiter and Neptune, and the Hildas asteroids are captured, as well as the irregular satellites of the giant planets\,; the Kuiper Belt is also shaped. The successes of the Nice model suggest that the planets were not formed where they now orbit, and the standard Minimum Mass Solar Nebula should be revised.

\section{Meteorites and Asteroids}
\label{met}

Meteorites are classified in 3 categories\,: the iron meteorites, the achondrites, and the chondrites. Iron meteorites are composed mainly of Iron and Nickel, and are considered as pieces of the metal core of a differentiated body. Achondrites are also called stony meteorites, and are considered as pieces of the mantle or the crust of a differentiated body. Chondrites are the most numerous, and considered as pieces of undifferentiated bodies\,; they are made of three components\,: the Calcium-Aluminum rich Inclusions (CAI), the chondrules, and the matrix which glues them together. The CAI are millimeter size grains made of refractory elements (mainly Calcium and Aluminum), that are supposed to condensate first in the protoplanetary disc. Therefore, the CAI formation traditionally marks time 0 of the solar system, $4.569\times 10^9$ years ago. The chondrules are millimeter spheres, about 2 million years older than the CAI (Amelin et al. 2002\,; Bizzarro et al. 2004) (see below how such a relative datation is possible). Chondrules constitute up to 80\% by volume of chondrites. As they are the main component of the most numerous and most primitive meteorites, chondrules are usually considered as the elementary bricks of planetesimals and planets.

\subsection{Short-lived radionucleides}

The relative datation of meteorites or pieces of meteorites (like the CAI and the chondrules) is generally done using short-lived, extinct radionucleides. For instance, $^{26}\!$Al decays in $^{26}$Mg with a half-life $\tau = 0.7$ million years. The isotopic ratio \mbox{$^{26}\!$Al/$^{27}\!$Al} in the solar nebula decreases therefore with $\exp(-t/\tau)$. When a body forms, it captures some Aluminum, with the isotopic ratio available at that time. Then, the $^{26}\!$Al decays in $^{26}$Mg {\it inside the body}. Now, all the $^{26}\!$Al has decayed, and the measured \mbox{$^{26}$Mg$/^{27}\!$Al} ratio reflects the initial \mbox{$^{26}\!$Al$/^{27}\!$Al} ratio. Therefore, the higher the present \mbox{$^{26}$Mg$/^{27}\!$Al} ratio in an Aluminum rich body, the earlier it formed.

The question is then\,: where did the short-lived radionucleides come from\,? A popular hypothesis for the origin of $^{26}\!$Al (and $^{60}$Fe) is a contamination caused by a near-by supernova. Gounelle \& Meibom (2007) have shown that in this case, the supernova should have also injected a considerable amount of $^{16}$O in the solar nebula, but not in the Sun because the Sun is too small to be contaminated. Therefore, the present isotopic ratio $^{17}$O$/^{16}$O should be smaller in the Solar System than in the Sun itself. This could be detected by the Genesis mission.

In 2008, the same authors have calculated the probability of such a supernova contamination for a protoplanetary disc in a young cluster (like the Orion nebula). They find that it is smaller than $3\times 10^{-3}$. Therefore, the supernova contamination scenario is very unlikely (but not impossible). They conclude that fluctuations of the interstellar medium composition are a more plausible explanation for the $^{60}$Fe content of the Solar System.

\subsection{Evidences for early differentiation}

The standard idea of terrestrial planet formation is that after the formation of chondrules, solid grains (CAI, chondrules, \ldots) gather, form little asteroids that collide, merge, and grow in a few million years. If they grow big enough, these planetary embryos can melt and differentiate. But a few recent results suggest that differentiated bodies existed already at the time of CAI formation.

Kleine et al. (2005a,b) have found that some iron meteorites are 2 to 3 million years older than the chondrules, using an other short lived radionucleide\,: $^{182}$Hf that decays in $^{182}$W in 9 million years. Therefore, they claim that ``{\it Tungsten isotopes provide evidence that core formation in some asteroids predates the formation of chondrite parent bodies}''. It seems that some iron meteorites formed even before the CAI. This can't be explained in the above standard scenario.

In addition, Bizzarro et al. (2005) found a $^{26}$Mg excess in eucrites and meso\-si\-de\-rites (other sorts of achondrites). The parent bodies of these meteorites must therefore have differentiated significantly before that $^{26}\!$Al was extinct, that is within $\sim 2$ million years after CAI formation.

Finally, Baker et al. (2005) have found basaltic achondrites (angrites) dated only 1 million years after the CAI. Thus, volcanism should have happened at that time, before the formation of chondrules.

In the end, there are strong and converging evidences, coming from various sorts of differentiated meteorites, that differentiation of solid bodies took place in the very beginning of the Solar System, before the chondrules appeared. This contradicts the standard model of planet formation in which planets form from chondrules over the time of a few million years, and they can differentiate only when they reach a critical size.

\subsection{Iron meteorites as remnants of planetesimals formed in the terrestrial planet region}

This subsection is named after the corresponding article by Bottke et al. (2006), in which the authors make N-body simulations of test particles placed initially within 2 AU in the presence of Moon to Mars sized planetary embryos. They find that the embryos scatter the test particles, and spread them into the Main Belt region. In less than 10 million years, 10\% of the particles initially located between $1.5$ and $2$ AU settle in the Main Belt. Even some of the ones originating from $0.5-1$ AU end in the Main Belt (up to 0.1\%). Then, these interlopers behave dynamically in the same way as the original Main Belt bodies.

In fact, collisions destroy 90\% of the bodies smaller than 100 km inside 1.5 AU. A differentiated small body would explode into iron meteorites and stone meteorites. These meteorites would then behave like the above test particles, and have a chance to reach the Main Belt (particularly its inner part). Then, the Yarkowski effect and the spatial weathering eliminates the fragile and poor thermally-conductive stony meteorites, and we are left with a large population of iron meteorites in the Main Belt, coming from different parent bodies. In this view, most of the iron meteorites now observed coming from the Main Belt are in reality pieces of small differentiated bodies that formed early in the terrestrial planet region, within 2 AU from the Sun. This explains that the parent bodies of the iron meteorites represent two thirds of the parent bodies of all meteorites (while iron meteorites are rare), and that they were small (20-200km). Note that this concerns only the population of the iron asteroids in the Main Belt. Most the asteroids now present in the Main Belt were formed there.

\subsection{Pieces of planetesimals as seeds for chondrules}

With electron microscopy, Libourel \& Krot (2007) have observed seven polished thin sections of the Vigarano meteorite (a CV chondrite). Inside some of the chondrules, they found ``{\it olivine-rich aggregates showing granoblastic textures and composed of coarse-grained forsteritic olivines and} Fe,Ni{\it -metal nodules}''. These aggregates appear in the observed sections as a pavement of polygonal olivine grains. The olivine-olivine or olivine-metal junctions between grains are either ``dry'' (without any glass), or ``wet'' (separated by thin layers of glass). The wet ones present wetting angles smaller than $60^\circ$. This indicates that the melt infiltrated inside a dry junction. If the grains had been initially separated and had joined by expelling the melt, the wetting angle would have been larger. Therefore, this texture ``{\it can only be produced by sintering and prolongated annealing of pre-existing material at high temperature}''. Shortly said, these aggregates existed before the chondrule, and they formed in a high pressure, high temperature environment.

The authors conclude that type~I chondrules formed from a tiny piece of the mantle of a pre-existing planetesimal. This millimeter-sized piece of rock was then melted and cooled down, accreting other components and acquiring a spheroidal shape.

\subsection{Conclusion}

All these results gather to build a consistent scenario of the first phase of planetary formation in the Solar System, that differs from the previous standard idea. A rapid formation of planetesimals during the first million years after CAI formation (or even before) seems very likely in the region of the terrestrial planets. Here, the dynamical timescale is indeed small, and allows for the processes of planet formation to happen in a short delay. As the short-lived radionucleides are not yet extinct, they produce heat by decaying, and this enables small bodies to melt and differentiate.

These small, early formed, differentiated bodies then suffer destructive collisions. The large mantle fragments are eroded and can not survive a few billion years\,; they are not observable anymore. Some of the cores fragments reach the Main Asteroid Belt, where they are now observed as iron meteorites. In the destructive collisions, tiny fragments of rock are also produced\,: they later transform into chondrules.

These chondrules can gather to form larger bodies, and planet formation continues, with these bricks. However, the short-lived radionucleides are extinct at that point, and therefore only the bodies larger than $\sim400$ km differentiate. Thus, undifferentiated asteroids and chondrites are observed today.


\section{Earth and Moon}
\label{EM}

After a runaway and oligarchic growth of embryos, the last stage of the formation of terrestrial planets is a phase of giant impacts between the Mars sized embryos. This phase lasts until a stable configuration such as the present one is found. The formation of the Moon is explained by such ``{\it a giant impact near the end of the Earth's formation}'' (Canup \& Asphaug, 2001).

\subsection{Equilibration}

The Earth and the Moon have undistinguishable Oxygen isotopic composition (Wiechert et al., 2001). In contrast, all other bodies of the Solar System have different Oxygen isotopic composition. To explain this, Pahlevan \& Stevenson (2007) have proposed a mechanism of equilibration. Just after the impact, the Earth is covered by a magma ocean, and surrounded by a disc of molten material. A silicate vapor atmosphere of about $2\times 10^{-3}$ Earth masses embraces the Earth and the disc. Turbulence in the liquids (disc and magma ocean), and exchanges between the liquid and gaseous phases make the entire system isotopically homogeneous. This works for Oxygen as well as other elements, like Hafnium.

After that, to keep the composition of both, the Earth and the Moon, unchanged, no significantly massive body should have collided with any of them. Thus, the Moon forming impact should be the last giant impact in the history of the Earth.

\subsection{Age of the Moon}

The age of the Moon can be estimated with the $^{182}$Hf/$^{182}$W chronometer. As mentioned above, $^{182}$Hf decays in $^{182}$W with a half-life of 9~Myrs. But Hafnium (Hf) is lithophile, while Tungsten (W) is siderophile. Therefore, during differentiation, the Hafnium goes in the mantle while the Tungsten gathers in the core. If the differentiation occurs when all the $^{182}$Hf has decayed into $^{182}$W, all the W reaches the core, and the mantle is $^{182}$W poor. If the differentiation is finished while $^{182}$Hf is still active, some $^{182}$Hf is caught by the mantle and decays there into $^{182}$W, leading to a $^{182}$W excess in the mantle. Therefore, the stronger the excess of $^{182}$W in the mantle, the earlier the differentiation took place.

As an excess of $^{182}$W is observed in the lunar mantle, Lee et al. (2002) and Kleine et al. (2005c) concluded that the Moon formed and differentiated within 60~Myrs, and probably even earlier.

\vspace{12pt}
N-body numerical simulations show that the giant impacts phase should last about 40~Myrs, if Jupiter and Saturn have their present eccentricities\,; this implies that the Moon is about 40 million years old, consistently with the above estimate.

However, if one assumes that at that time, Jupiter and Saturn were on circular orbits, one finds that this phase lasts about 100~Myrs (O'Brien et al., 2006). A circular initial orbit for Jupiter and Saturn is favored for a few reasons. First, it is required in the Nice model (see section~\ref{Nice}). Second, O'Brien et al. (2006) have shown that this hypothesis enables the terrestrial planets to accrete about 15\% of water-rich material from the outer Main Asteroid Belt, which explains the presence of water on the terrestrial planets today. Third, circular orbits of Jupiter and Saturn enable embryos to survive longer in the Main Belt, which explains better the properties of the Asteroid Belt (O'Brien et al. 2007\,; Petit et al. 1999). This would lead to an age of $\sim 100$~Myrs for the Moon, in contrast with the result of the Hf/W chronometer.

\vspace{12pt}
Yet, Touboul et al. (2007) have shown that ``{\it the dominant $^{182}$W component in most lunar rocks reflects cosmogenic production}''. The $^{182}$W excess measured is in fact pollution due to cosmic rays. New data from unpolluted samples (KREEP) lead the authors to conclude that ``{\it lunar and terrestrial mantles have identical $^{182}$W/$^{184}$W. This constrains the age of the Moon and Earth to 62~Myrs (+90/-10)}''. This result has been confirmed by the same authors in 2008 on another lunar sample\,: plagioclase separates from two ferroan anorthosites (pieces of the crust). In addition, they measured the basalts cosmogenic pollution.

\vspace{12pt}
Finally, the N-body simulations of the giant impacts phase and the $^{182}$Hf/$^{182}$W chronometer are consistent again. The revised age of the Moon is $\sim 100$~Myrs.

Combining the results of the N-body simulations and the new measures of the age of the Moon via the $^{182}$Hf/$^{182}$W chronometer, one can also conclude that the isotopic datation favors the hypothesis that Jupiter and Saturn were on circular orbits in the first hundred million years of the Solar System.

\subsection{Re-equilibration during collisions}

When two differentiated bodies collide, their mantles may have different $^{182}$W excesses. If during the collisions, the cores merge and the mantles merge without core-mantle interaction, the $^{182}$W excess in the mantle of the final body is the average of the two initial $^{182}$W excesses. But if the core of the impactor is disrupted and the metal sinks in centimetre-sized drops through the mantle of the target, it takes all the Tungsten with him down to the core. Then, the $^{182}$W excess in the mantle is reset. In that case, the measured $^{182}$W excess does not date the differentiation but the impact.

Nimmo \& Agnor (2006) have studied the ``{\it isotopic outcome of N-body accretion simulations}''. They find that a better match with the observations of the planets and asteroids can be achieved if one assumes many giant impacts after 10~Myrs, with a complete re-equilibration.

\subsection{Summary}

The Moon forming impact occurred at the end of the giant impacts phase of the formation of the terrestrial planets. That is most likely between 60 and 100 million years after Solar System formation.

During the Moon formation, there was an isotopic equilibration process in Oxygen and other elements. But the $^{182}$Hf radioactivity was extinct, so that the Moon and the Earth mantles still have the same $^{182}$W/$^{184}$W ratio.

\section{The Nice model and its applications}
\label{Nice}

\subsection{Migration in the Solar System}
\label{migr}

It is well known that planets in protoplanetary gaseous discs migrate. In isothermal discs, low mass planets suffer inwards type~I migration, at a rate proportional to their masses. Giant planets open gaps, and then follow the global viscous evolution of the disc, generally accretion towards the star\,; this is called type~II migration. In this framework, the reason why the giant planets of the Solar System haven't come close to the Sun has been a longstanding mystery.

Masset \& Snellgrove (2001) have shown that if Jupiter and Saturn orbit in mean motion resonance in a common gap, the pair of planet decouples from the disc evolution and may migrate outwards. This is because Saturn is less massive than Jupiter\,: therefore the force exerted by the disc on the innermost planet is larger than the one on the outer planet, and the pair of planets is not in equilibrium in the gap. Pierens \& Nelson (2008) have shown that the capture in the 3:2 mean motion resonance is the most likely outcome for the Jupiter-Saturn pair in a disc. Finally, Morbidelli \& Crida (2007) studied the influence of the disc properties on the migration rate of the Jupiter-Saturn pair in 3:2 resonance, and found that their migration is stopped in some cases. Those solutions allow Jupiter and Saturn to form in the outer Solar System and to stay there, without migrating in the region of the terrestrial planets.

From the stationary configuration of Jupiter and Saturn in the disc found by Morbidelli \& Crida (2007), Morbidelli et al. (2007) added Uranus and Neptune to the system. The ice giants migrate inwards and are then captured in mean motion resonance with Saturn or Uranus. A prevention of the migration of the four giant planets of the Solar System is therefore possible, if they are in a compact configuration, very different from the present one. Six possible, fully resonant configurations have been found. They are admittedly not similar to the present structure of the outer Solar System, but the authors have shown that two of them are compatible with the Nice model.

\subsection{The Nice model}

In a trilogy of Nature papers published in 2005, Gomes, Tsiganis, Morbidelli, \& Levison proposed a new model to explain at the same time the architecture of the outer Solar System (Tsiganis et al., 2005) and the Late Heavy Bombardment (Gomes et al., 2005). In this model --\,often referred to as the ``Nice model'', because its four authors were working at Observatoire de la C\^ote d'Azur in Nice\,-- the evolution of the planets is followed \emph{after} the gas disc has dissipated. The four giant planets are orbiting at first on circular orbits, in a very compact configuration\,: the radius of the orbit of Jupiter is 5.45 AU, that of Saturn is 8.45 AU, Uranus is between 11 and 13 AU, while Neptune is between 14 and 17 AU. A disc of planetesimals extends outside of the orbit of the last planet and up to about 35 AU\,; its total mass is about 35 Earth masses.

By scattering planetesimals, the planets slowly change their orbits. On average, Saturn, Uranus, and Neptune move outwards, while Jupiter moves inwards. At some point (after a few hundred million years), Jupiter and Saturn cross their 2:1 mean motion resonance. This increases their eccentricities, which destabilises the entire system. Uranus and Neptune have close encounters with each other and with Saturn, and they enter the disc of planetesimals, which is destroyed as a consequence. Most of the planetesimals are scattered by the planets and some of them reach the inner solar System, producing the Late Heavy Bombardment. The amount of planetesimals that hit the Moon is in agreement with the estimates of the LHB. In about a hundred million years, the planets reach their present semi major axes, and their eccentricities are damped by dynamical friction with the planetesimals. The average outcome of the global instability matches very well the present position of the planets, for a reasonable range of parameters.

In addition to the Late Heavy Bombardment and the present orbital parameters of the giant planets, a few characteristics of the Solar System can be explained in the frame of the Nice model, thanks to the global instability. These positive results will be reviewed in the following subsections.

\subsubsection{Trojans and Hildas capture}

Morbidelli et al. (2005) show that during the global instability, the co-orbital region of Jupiter becomes chaotic. It is dynamically open, in contrast to the present situation where it is closed. Therefore, any Trojan asteroid that would have been present before the global instability should leave, but planetesimals passing by can enter this zone. As a lot of bodies coming from the outer planetesimal disc cross Jupiter's orbit during the global instability, the co-orbital zone of Jupiter is occupied by a transient population\,: planetesimals enter and leave this region. When the instability is over, the chaos disappears and the region is dynamically closed again. The planetesimals present at this time are captured forever, and become the Trojans that are observed today. The distribution of the Jupiter Trojans in eccentricity, inclination, and libration angle can be explained by this process. In contrast, their large inclinations could not be understood before.

A similar process also explains the capture of the Trojans of Neptune (Nesvorn\'y \& Vokrouhlick\'y, personal communication).

Bottke et al. (2008) find that planetesimals are also captured in the region of the Hildas asteroids, and in the outer Main Belt. They study their collisional evolution over 3.9 billion years, that is from the Late Heavy Bombardment until now. In the outer Main Belt, 90\% of the captured objects are eliminated. The final size distribution of the Hildas and of the Trojans is in excellent agreement with observations.

This suggests that the asteroids of spectral type D, that constitute the Trojans, the Hildas, and are also present in the outer part of the Main Asteroid Belt, are captured comets. They are remnants of the disc of planetesimals that used to exist beyond the orbit of Neptune.

\subsubsection{Irregular satellites}

During the global instability, Saturn, Uranus, and Neptune have close encounters with each other, while the density of planetesimals in the region of the giant planets is high. Therefore, encounters involving two planets and at least one planetesimal should happen. In that case, it is possible that a planet captures a planetesimal. This body becomes a satellite of the planet, but there is no reason why it should orbit in the equatorial plane of the planet. Therefore, it is an irregular satellite.

Nesvorn\'y et al. (2007) show that this mechanism may well explain the irregular satellites observed for Saturn, Uranus, and Neptune (Jupiter should have acquired its own by an other process). Indeed, they find that any planetesimal has a capture probability by an ice giant of a few $10^{-7}$. The distribution of orbits of the captured satellites in inclination, eccentricity, and semi major axis, is in good agreement with observations for all the three planets. However, the size distribution should be evolved to match the observations correctly\,; this may be possible by collisional processes during the last 4 billion years.

\subsubsection{Kuiper Belt}

The disc of planetesimals lost about $99\%$ of its mass during the global instability. The remaining bodies represent the Kuiper Belt. Levison et al. (2008) have shown that the main properties of the Kuiper Belt can be explained in this frame. Indeed, while Neptune has a large eccentricity after some close encounters with Uranus or Saturn, its outer Lindblad resonances overlap. The region between Neptune and its 2:1 mean motion resonance becomes therefore chaotic, and the planetesimals can travel through it, and fill this zone. Once Neptune's eccentricity is damped, the dynamical state of the region is frozen.

Under some reasonable assumptions, the simulations performed by the authors of the evolution of the planetesimal disc during and after the global instability reproduce the main observed properties of the Kuiper Belt\,:
\begin{enumerate}
\item the co-existence of resonant and non-resonant populations,
\item the peculiar distribution of the classical belt in semi major axes and eccentricities,
\item the existence of an outer edge at the location of the 2:1 resonance with Neptune,
\item the bi-modal inclination distribution of the classical population, with correlations between physical properties and inclination,
\item the orbital distribution of the Plutinos and the 2:5 librators,
\item the existence of the extended scattered disc,
\item the mass deficit.
\end{enumerate}

All these intriguing properties had never been explained all together in a single model. This successful reproduction of the Kuiper Belt is therefore a strong, new argument in favor of the Nice model.

\subsubsection{A new Minimum Mass Solar Nebula}

If one believes that the Nice model is true, it is clear that the four giant planets were not at their present position after the gas disc phase. Therefore, the standard Minimum Mass Solar Nebula (MMSN) is out of date. Desch (2007) assumed that the giants formed at their initial positions in the Nice model (see above), exchanging Uranus and Neptune (which happens in $50\%$ of the cases in the original Nice model simulations). He finds that the surface density of the protoplanetary disc needed to build the planets and the outer disc of planetesimals, can be very well fitted by a steep power law\,: $\Sigma \approx 50500 (r/1{\rm AU})^{-2.168}$ g.cm$^{-2}$. This density is one order of magnitude larger at 5~AU than the Hayashi (1981) MMSN\,: $\Sigma \approx 1700 (r/1{\rm AU})^{-1.5}$ g.cm$^{-2}$.

Inquiring the time evolution of the Solar Nebula, Desch finds a solution of a decretion disc, photoevaporated at 61~AU, fed by the internal parts. The density profile could remain almost unchanged for about ten million years. Therefore, the solid cores of the giant planets have time to reach their isolation masses, and to accrete gas. The possible formation of the giant planets with their masses and at the positions requested for the Nice model make this new MMSN attractive.

However, the planetary migration in this dense disc has not been questioned. The gas density at the location of Jupiter is so high that this planet should enter in type~III, runaway migration towards the Sun, and disappear in a few hundred years. The mechanism of Masset \& Snellgrove (2001) used by Morbidelli \& Crida (2007) and Morbidelli et al. (2007) (section~\ref{migr}) can not work in this disc. The four giants are inevitably lost in less than 20000 years (see Crida, 2009) ; this is a severe issue. On the contrary, planetary migration could account for a formation of the giant planets on a larger radial range, in a less dense disc, followed by a compactification of the configuration by migration. Therefore, I would claim that a new Solar Nebula consistent with the Nice model is still to be built.

\subsection{Conclusion on early dynamics of the giant planets}

The planets did not form where they are now observed. First, they migrated inside the protoplanetary gaseous disc. Their migration all the way down to the Sun can be prevented by resonances. Then, their orbital parameters probably dramatically changed about 700 millions of years after the protoplanetary disc dissipation, through a global dynamical instability.

The Nice model assumes a compact configuration of the giant planets after the gas disc phase\,; this configuration is slowly perturbed by an outer disc of planetesimals\,; then, the system is destabilised by the 2:1 resonance crossing of Jupiter and Saturn, producing the Late Heavy Bombardment and driving the planets on their present orbits.

In the frame of this model, the orbital and size distribution of the Trojan and Hildas asteroids can be explained. The structure of the Kuiper Belt is also well reproduced. In addition, the irregular satellites of Saturn, Uranus and Neptune can be captured during the global instability. For these reasons, the Nice model is one of the most impressive results of the last decade on Solar System formation.

\section{Conclusion}

In the last three years, a few little revolutions took place in planetary science. Planetesimal formation and differentiation now seems to predate the chondrules. The age of the Moon forming impact has been revised to more than 60 Myrs. And finally, the Nice model shows that after the gas disc dissipated, the Solar System has not yet its final aspect at all\,; on the contrary, a late global instability shapes its present architecture.

These new ideas are promoted by several results, so that they are robust and now accepted by most researchers. They offer the possibility to draw a consistent history of the formation of the Solar System. Let us summarise this new scenario. First, planet formation takes place in less than a million years in the innermost regions. The formed bodies differentiate thanks to the heat produced by the decay of short-lived radionucleides. The differentiated bodies are then destroyed by collisions. This produces the seeds for the formation of chondrules, and provides iron meteorites that will be scattered into the Main Asteroid Belt. Then, planetary formation continues in the protoplanetary disc. Jupiter, Saturn, Uranus and Neptune avoid migration towards the Sun by adopting a compact configuration, on circular orbits. Then, the gaseous proto-planetary disc disappears, and the last phase of the formation of the terrestrial planets (giant impacts) lasts about a hundred million years. In the end comes the Moon forming impact on the Earth, while the $^{182}$Hf radioactivity is already extinct. Complete isotopic equilibration between the molten Earth and the molten disc precursor of the Moon takes place through a common silicate atmosphere. About 600 million years later, a global instability in the dynamics of the giant planets causes the Late Heavy Bombardment of the terrestrial planets, drives the giant planets on their present orbits, enables them to capture their Trojans asteroids and irregular satellites, and shapes the Kuiper Belt.

This scenario would have been unrealistic science-fiction in 2003. It looks in 2008 beautiful and consistent. The last five years have been fascinating for Solar System science. There are still a few mysteries to explain, and the consequences of these major changes in the chronology of the formation of the Solar System have not yet been all explored. This opens the possibility for a new exciting decade.

\

\subsection*{Acknowledgments}

I wish to thank the conveners of the Symposium~3 ``Planetary Formation and Extra-solar planets'', A. {\sc Dutrey}, W. {\sc Kley}, and E. {\sc Pilat-Lohinger}, for having invited me to give this review talk, and given me the opportunity to present all these exciting results at the Joined European and National Astronomy Meeting 2008.\\
F. {\sc Nimmo}, A. {\sc Morbidelli}, and T. {\sc Kleine} are also acknowledged for providing me with several graphs to illustrate my presentation, as well as Ines and Ameline for their help in the redaction of this proceeding.

\subsection*{References}

{\small
\bref
Amelin, Y., Krot, A.~N., Hutcheon, I.~D., \& Ulyanov, A.~A.  2002, Science, 297, 1678

\bref
Baker, J., Bizzarro, M., Wittig, N., Connelly, J., \& Haack, H.  2005, Nature, 436, 1127

\bref
Bizzarro, M. Baker, J.~A., \& Haack, H.  2004, Nature, 431, 275

\bref
Bizzarro, M. Baker, J.~A.,  Haack, H., \& Lundgaard, K.~L.  2005, ApJ, 632, L41

\bref
Bottke, W.~F., Nesvorn\'y, D., Grimm, R.~E., Morbidelli, A., \& O'Brien, D.~P.  2006, Nature, 439, 821

\bref
Bottke, W.~F., Levison, H.~F., Morbidelli, A., \& Tsiganis, K.,  2008, in Lunar and Planetary Institute Conference Abstracts, Vol. 39, Lunar and Planetary Institute Conference Abstracts, 1447--+

\bref
Canup, R.~M. \& Asphaug, E.  2001, Nature, 412, 708

\bref
Crida, A.  2009, ApJ, submitted

\bref
Desch, S.~J.  2007, ApJ, 671, 878

\bref
Gomes, R., Levison, H.~F., Tsiganis, K., \& Morbidelli, A.  2005, Nature, 435, 466

\bref
Gounelle, M. \& Meibom, A.  2007, ApJ, 664, L123

\bref
Gounelle, M. \& Meibom, A.  2008, ApJ, 680, 781

\bref
Hayashi, C.  1981, PThPS, 70, 35

\bref
Kleine, T., Mezger, K., Palme, H., \& Scherer, E.  2005a, in Lunar and Planetary Institute Conference Abstracts, Vol. 36, 36th Annual Lunar and Planetary Science Conference, ed. S. Mackwell \& E. Stransbery, 1431--+

\bref
Kleine, T., Mezger, K., Palme, H., Scherer, E. \& M\"unker, C.  2005b, Geochim. Cosmochim. Acta, 69, 5805

\bref
Kleine, T., Palme, H., Mezger, K., \& Halliday, A.~N.  2005c, Science, 310, 1671

\bref
Lee, D.-C., Halliday, A.~N., Leya, I. Wieler, R., \& Wiechert, U.  2002, Earth and Planetary Science Letters, 198, 267

\bref
Levison, H.~F., Morbidelli, A., Vanlaerhoven, C., Gomes, R., \& Tsiganis, K.  2008, Icarus, 196, 258

\bref
Libourel, G., \& Krot, A.~N.  2007, Earth and Planetary Science Letters, 254, 1

\bref
Masset, F. \& Snellgrove, M.  2001, MNRAS, 320, L55

\bref
Morbidelli, A. \& Crida, A.  2007, Icarus, 191, 158

\bref
Morbidelli, A., Levison, H.~F., Tsiganis, K., Gomes, R.  2005, Nature, 435, 462

\bref
Morbidelli, A., Tsiganis, K., Crida, A., Levison, H.~F., Gomes, R.  2007, AJ, 134, 1790

\bref
Nesvorn\'y, D., Vokrouhlick\'y, D., \& Morbidelli, A.  2007, AJ, 133, 1962

\bref
Nimmo, F., \& Agnor, C.~B.  2006, Earth and Planetary Science Letters, 243, 26

\bref
O'Brien, D.~P., Morbidelli, A., \& Levison, H.~F.  2006, Icarus, 184, 39

\bref
O'Brien, D.~P., Morbidelli, A., \& Bottke, W.~F  2007, Icarus, 191, 434

\bref
Pahlevan, K. \& Stevenson, D.~J.  2007, Earth and Planetary Science Letters, 262, 438

\bref
Pierens, A. \& Nelson, R.~P.  2008, A\&A, 482, 333

\bref
Touboul, M., Kleine, T., Bourdon, B., Palme, H., \& Wieler, R.  2007, Nature, 450, 1206

\bref
Touboul, M., Kleine, T., Bourdon, B., Palme, H., \& Wieler, R.  2008, in Lunar and Planetary Institute Conference Abstracts, Vol. 39, Lunar and Planetary Institute Conference Abstracts, 1940--+

\bref
Tsiganis, K., Gomes, R., Morbidelli, A., \& Levison, H.~F.  2005, Nature, 435, 459

\bref
Wiechert, U., Halliday, A.~N., Lee D.-C., et al.  2001, Science, 294, 345
}

\vfill

\end{document}